**Terahertz-permittivity of Carbon Nitrides: Revealing humidity-enhanced dielectric properties on the picosecond timescales relevant for charge carrier photogeneration**

**Authors:**

Reehab Jahangir[1,2,§], Filip Podjaski[1,§,*], Paransa Alimard[1], Sam A. J. Hillman[1], Stuart Davidson[2], Stefan Stoica[2], Andreas Kafizas[1], Mira Naftaly[2], James R. Durrant[1]

[1]Department of Chemistry and Centre for Processable Electronics, Imperial College London, UK

[2]National Physical Laboratory, Teddington, UK

[§]equally contributed

*Correspondence to: f.podjaski@imperial.ac.uk

# Abstract

Organic based semiconductor materials offer emerging and sustainable solutions for solar energy conversion technologies and electronics. However, knowledge of their intrinsic (photo)physical properties and light-matter interactions is often limited, especially with respect to the frequency dependent dielectric properties on the relevant timescales of exciton separation and charge generation (fs-ps). By using terahertz-time-domain spectroscopy (THz-TDS), we show that the complex permittivity and THz conductivity of different polymer materials and graphitic carbon nitrides - melon and poly(heptazine imides) (K-PHI) - can be measured directly and accurately in different environmental humidities. Its effects are most strongly observed in the ionic and 2D carbon nitride, K-PHI, where both real permittivity and THz conductivity double from dry to humid conditions (~4-8 and 75-150 S/m, respectively) surpassing the intrinsic dielectric response from water or K-PHI through synergistic effects. Our findings are backed by fs-ps transient absorption spectroscopy (TAS), confirming the impact of humidity on light conversion behaviour on the ps-timescale in K-PHI. We highlight the importance of dielectric property characterization at THz frequencies as critical for understanding photophysical behaviour at exciton or charge separation time scales, especially in the presence of water and hydrated ions, which may also be beneficial for computation, exploring next generation photocatalysts, electronics or ionotronics.

**Keywords:** Carbon nitrides, Terahertz, complex permittivity, dielectric properties, solar energy conversion

# Broader Context

Solar energy conversion technologies are widely explored for renewable energy supply in form of electricity and fuels, for sensors etc. Most rely on photovoltaic effects, where light absorption generates excitons that are subsequently separated into functional charges. This transition must overcome the exciton binding energy which is inversely proportional to the materials' dielectric response $\varepsilon_r$. Its highly frequency dependent value is typically unknown in the corresponding fs-ps regime (THz gap). Hence, it is approximated from static or low frequency values, or those at much higher frequency. Using THz time domain spectroscopy, we show that the full complex permittivity, i.e. the dielectric response and loss factors related to THz conductivity, can be measured directly and non-invasively, on the corresponding time scales of exciton separation and charge generation. Our study focussed on various polymers and Carbon Nitrides ($CN_x$), and the influence of environmental effects in wetting materials. It reveals that in ambient conditions, carbon nitrides have unexpectedly high THz dielectric values in the range of oxide semiconductors. Especially in the ionic $CN_x$ K-PHI, the dielectric response can be increased by >100% ($\varepsilon_r$ >8) through synergistic humidity effects, giving also rise to high THz conductivity. This highlights the crucial importance of permittivity probing in relevant conditions.

# Introduction

Organic semiconductor materials are increasingly used for optoelectronic and energy conversion applications since they are often easily processable and offer bottom-up tunability for desired properties and functions[1]. A potential pathway for their exploitation in a sustainable manner, addressing increasing environmental challenges, lies in their ability to absorb visible light and to utilise this energy for electricity generation in solar cells, as photodetectors, or for the direct photocatalytic production of fuels and chemicals. In all these processes, the absorption of photons typically generates bound excitons, that are separated into charges, driving the desired function. However, in many organic based semiconductors, such as carbon nitrides ($CN_x$), covalent organic frameworks, or polymer systems, light energy conversion is still less efficient than in inorganic materials or the much-studied hybrid perovskites. In particular, the generation of charges from excitons in these materials is often a key challenge for efficient device performance[2,3]. The fundamental reason for the strong exciton recombination often observed in organic based semiconductors is linked to their relatively high exciton binding energy $\Delta E_{Ex\_BE}$ (compared with inorganic semiconductors), which is related to their intrinsic real permittivity $\varepsilon_r$ [4]:

$$\Delta E_{Ex\_BE} \approx \frac{e^2}{4\pi\varepsilon_o\varepsilon_r r} \qquad (Equation\ 1)$$

where $e$ is the electron charge, $\varepsilon_0$ is vacuum permittivity, $r$ is Coulomb radius[5].

With low $\varepsilon_r$, the Coulomb attraction of excitons is poorly screened by the materials' backbones, which affects the excitons' ability to dissociate, and thus hinders the generation, stabilization, and extraction of useful charges[6].

The complex permittivity $\boldsymbol{\varepsilon} = \varepsilon' + i\varepsilon''$ of materials, from which static real permittivity (also known as dielectric constant $\varepsilon_r = \varepsilon'$) is typically deduced in the low frequency limit, is a highly frequency dependent property, as depicted in **Figure 1a**. Crucially, often little is known about the complex permittivity properties of materials on the relevant timescale for exciton generation and separation - this typically being in the fs-ps regime, corresponding to THz frequencies. A further challenge is that the influence of environmental factors such as water exposure on the complex permittivity of the material is typically disregarded, with exciton separation often assumed to be an intrinsic material property. This environmental factor, however, is especially relevant in wetting or porous materials that have high volume interactions with their environment, and hence also in applications such as photocatalytic water splitting [7].

Dielectric properties are typically probed from the mHz up to the MHz regime by electrochemical impedance spectroscopy (EIS), with free space microwave dielectric measurements with Vector Network Analyzers (~0.5 – 300 GHz) [8–11], and by optical means (e.g. Spectroscopic Ellipsometry (SE)) at $10^{14}$-$10^{15}$ Hz (Peta-Hz), hence leaving out the important THz timescales - the so-called *THz-gap* (**Figure 1a**) [12–14]. Water has a very high DC real permittivity ($\varepsilon'_{H2O} \approx 74$), which has been shown to affect the intrinsic charge generation and stabilization of semiconductors[15]. However, at THz regimes, its value is very moderate ($\varepsilon'_{H2O,\ 1.5\ THz} \approx 4.15$) [16]. Both studies of the THz dielectric response of organic semiconductors, and the impact of water exposure on their response, have been very limited to date. Our study shows how to access the dielectric response of various organic semiconductors and its modification by humidity exposure, on time scales relevant for solar energy conversion and photocatalysis, by using terahertz time-domain spectroscopy (THz-TDS).

THz-TDS is an increasingly commonly available technique that allows to probe the complex permittivity **ε** of materials in their bulk (averaged over the volume of the probing beam interacting with the sample, no approximations required) and in different environments, thus providing important insights into their bulk photophysical properties on the ps timescale and below (1 THz ~1 ps) (**Figure 1a**)[17,18]. For this study, we chose different insulating polymers and the broadly employed organic semiconductor poly(3-hexylthiophene) = P3HT as references[19], while focussing on heptazine-based $CN_x$ materials, which are among the most widely used model organic semiconductors for various light driven energy conversion applications[4]. $CN_x$ are easy to synthesise and to post-functionalise, highly stable, and have visible light bandgaps (typically 2.7 to 3 eV) that are suitably positioned to drive various photo-redox reactions, such as photocatalytic water splitting[20]. Their application areas also include environmental remediation, photodetectors and sensors, ion pumps, light driven microswimmers, photocharging and dark photocatalysis, as well as photobatteries. The latter are often linked to ionic interactions with photogenerated charges[20].

We now present the first THz complex permittivity measurements on these layered materials, comparing different 1D melon-type $CN_x$, and 2D-linked poly(heptazine imide) PHI, which can contain different solvated ions in their pores that are affect their function and performance (see **Figure 1b**)[21–24]. We further study the influence of exposure to humidity (water), which commonly has a strong influence on ionic mobility, and show its potential to synergistically affect a material's real and imaginary permittivity at THz frequencies, in cases where water is strongly adsorbed[25–29]. The influence of an increase in intrinsic real permittivity and THz conductivity we describe herein is mirrored by more pronounced pre-charge signals of PHI in humid conditions, indicating modified exciton-to-charge relaxation mechanisms, measured by transient absorption measured by transient absorption spectroscopy (TAS) in the fs-ps regime, underlining the important influence of humidity on intrinsic photophysical properties. We discuss the findings in detail and especially in the context of photocatalytic reactions in water.

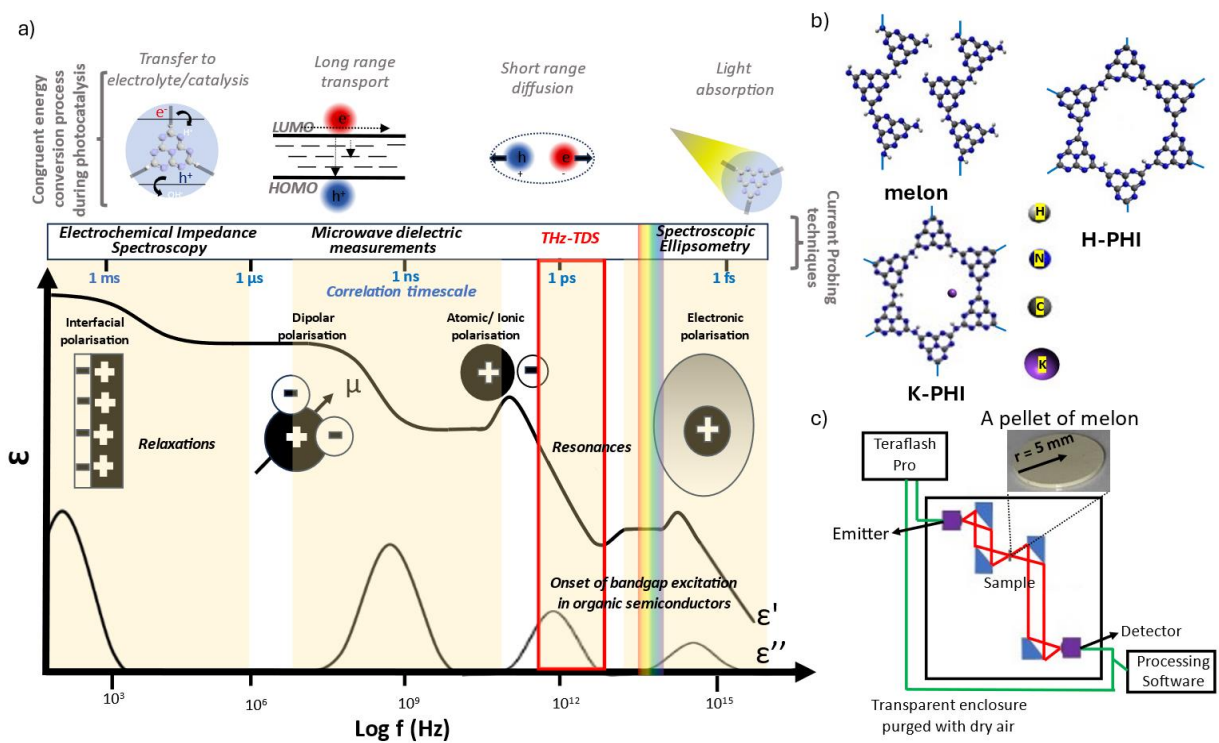

**Figure 1**: Frequency-dependent complex permittivity and $CN_x$ materials investigated. **a)** Different polarisation mechanisms are activated with increasing frequency of an electric field *E*, leading to frequency dependent (dispersive) complex permittivity **ε** responses. The top scale shows the correlation of the timescale to the frequency at which these polarisation events may influence (opto)electronic properties (reproduced from[30] under the CC BY 4.0 license). **b)** Schematics of atomic structures of the carbon nitrides investigated: the one-dimensional graphitic $CN_x$ *melon*[31], and the 2D poly(heptazine imide) PHI (here with $K^+$ and $H^+$ ions as K-PHI and H-PHI, respectively)[23]. **c)** Schematic of the transmission-mode terahertz time-domain spectroscopy (THz-TDS) set-up used and pictured, a processed sample of a melon pellet used for the study.

## Results

**Materials and Preparation:** The different $CN_x$ were synthesized following previously reported procedures (see *Experimental methods* for details). Three batches of melon-type $CN_x$ were synthesised at slightly different conditions (Melon # 1 by annealing urea at 550 °C in air followed by an acid/base wash; Melon #2 by annealing urea in air at 500 °C; and Melon #3 by annealing urea in nitrogen at 550 °C); two batches of K-PHI (derived from melamine and KSCN, K-PHI #1 by annealing melon in argon at 500 °C; and K-PHI #2 by annealing melon in air at 550 °C); two batches of the protonated form of PHI (H-PHI #1, #2) were obtained from the K-PHI's by acid washing[23]. Powder XRD data (**Figure S1**) shows characteristic peaks at 2Θ = 13.1° and 27.8° for melon, each corresponding to (100) and (002) the periodic arrangement of the tri-s-triazine rings and the interlayer stacking planes respectively[32–34], and the prominent ~10° and ~28° peaks for K-PHI and H-PHI, corresponding to the (010) and (001) planes respectively[24]. The optical band gaps, extracted from Tauc plots and diffuse reflectance measurements, are of approx. 3 eV for the melon samples and H-PHI, and ~2.8 eV for the K-PHI samples (**Figure S2**), as reported.

To study their complex permittivity by THz-TDS (see **Figure 1c**), thin pellets (d = 10mm) were pressed from the materials (see **Supplementary Information**, **Tables S1, S2** and **S3**). Since THz probes the entire sample volume, precise knowledge of porosity is required[35]. For this, the material density was measured by a He-pycnometer, resulting in 1.88(1) g/cm$^3$ for melon #1 and K-PHI in ambient conditions, and 1.85(1) g/cm$^3$ for H-PHI. Upon drying using conventional methods such as convection oven and desiccator, the amount of water adsorbed was reduced and stable densities of 1.72(1), 1.84(1), and 1.81(1) g/cm$^3$ were obtained respectively (see *Experimental Methods* for more details). These were subsequently used to calculate an effective pellet porosity (see **Figure S3** and **Supplementary Note 1**).

THz-TDS data was collected directly from these pellets in a custom-built chamber purged with dry compressed air to avoid the parasitic absorption of ambient water vapour (**Figure 1c**). The terahertz pulse transmitted through the sample is Fourier-transformed from the time to frequency domain. The measured change in phase and amplitude relative to the reference beam allows calculation of the complex refractive index via $\boldsymbol{n}$. The complex permittivity $\boldsymbol{\varepsilon}$ is extracted via $\boldsymbol{n} = \sqrt{\boldsymbol{\varepsilon}}$, including error propagation (**Supplementary Note 2**). $\varepsilon'(\omega)$ depicts the real permittivity (= dielectric polarisation reponse), and $\varepsilon''(\omega)$ the imaginary permittivity, which is a measure of energy dissipation due to loss mechanisms arising from polarisation, phonon contributions and acceleration of charge carriers due to the alternating electric field, relating linearly to THz conductivity ($\sigma_{THz}$, **Equation S11**).

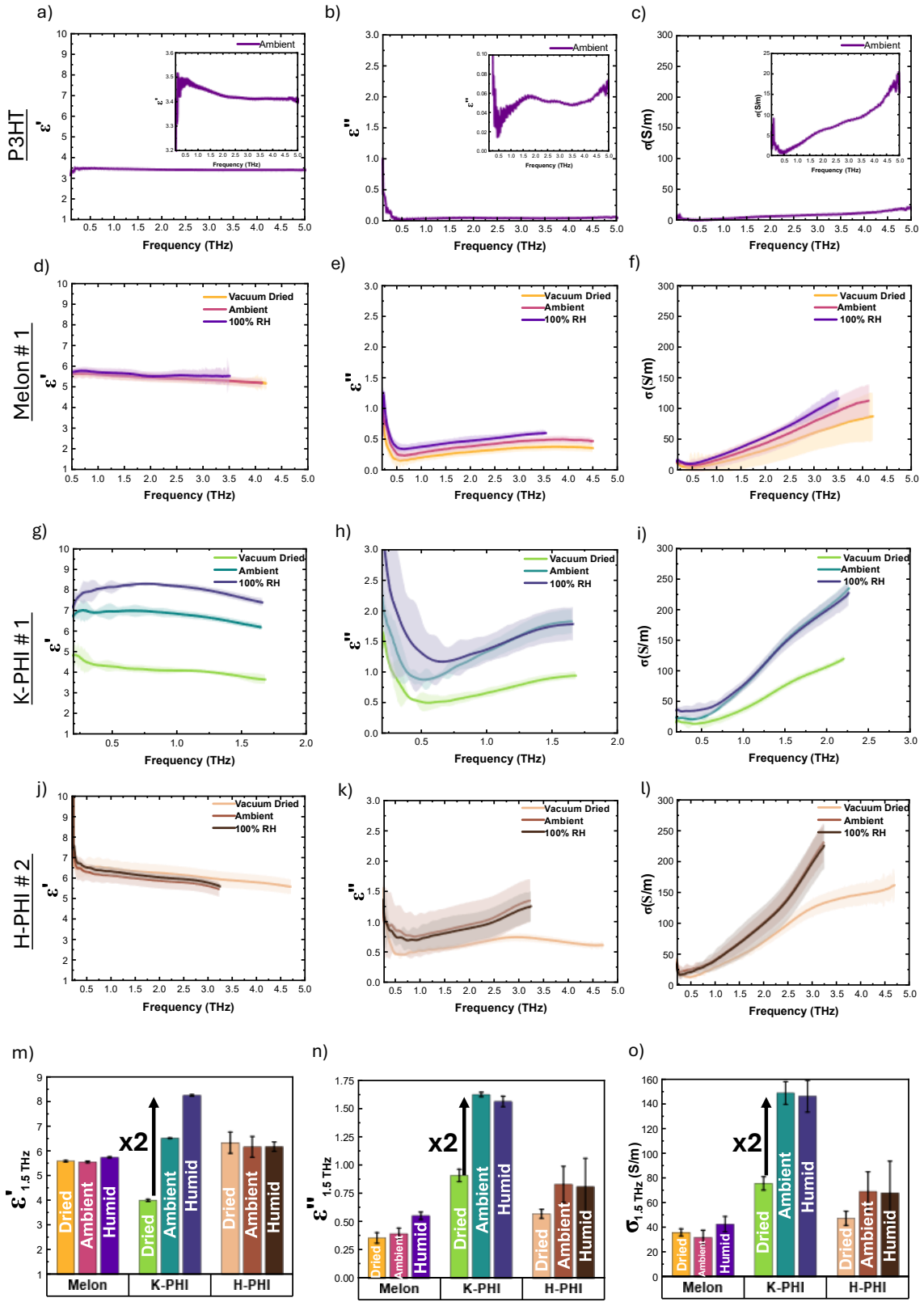

**Figure 2**: Real Permittivity (ε'), Imaginary Permittivity (ε'') and THz conductivity (σ) spectra of P3HT **(a,b,c)**, melon # 1 **(d,e,f)**, K-PHI # 1 **(g,h,i)**, H-PHI #2 **(j,k,l)** measured in vacuum dried, ambient and 100% RH conditions, respectively. **m,n,o**: Overview comparing the three $CN_x$ materials at 1.5 THz across different humidity environments.

**Ambient measurements:** First, data acquisition was veryfied on high density polyethylene (HDPE) and hygroscopic poly(ethyleneglycol) = PEG, confirming literature values and expected dispersion trends (**Figure S4, Supplementary Note 3**). Their real permittivity at 1.5 THz is $\varepsilon'_{1.5\,THz}$ = 2.39 and 2.76, respectively (**Table S4**). New measurements on the semiconductor P3HT (**Figure 2 a-c**) show a flat dispersion profile in the range probed, with $\varepsilon'_{1.5\,THz}$ = 3.44 and $\varepsilon''_{1.5\,THz}$ = 0.052 (**Supplementary Note 3**). All melon samples show very similar complex permittivity dispersion (**Figure S4, Supplementary Note 4**). Their $\varepsilon'_{1.5\,THz}$ = 5.5 is significantly higher than our reference materials (+57% vs. P3HT), and even water ($\varepsilon'$~4.15). The average $\varepsilon'$ of K-PHI is 6.75, which is 22% higher than the average melon value, with a less flat dispersion profile compared to all other materials. H-PHI has $\varepsilon'_{1.5\,THz}$ =6.51 (similar to K-PHI), but with 40% reduced $\varepsilon''_{1.5\,THz}$ (see **Table S4, Supplementary Note 3**).

Overall, from **Figure 2** and **S4**, one can see that the 2D PHI based materials have increased ambient real and complex permittivity values compared to 1D melon, and that the presence of alkali metal ions in the pores (here, $K^+$) of PHI seems to affect the dispersion profile of the complex permittivity spectra. Since these ions are not bound to the material backbone [23,24], it appears that they are freely moving and contributing to the dielectric response, perhaps with a different (saturating) behavior than the PHI$^-$ backbone at frequencies above 1.5 THz, giving rise to stronger levelling-off (decreasing) for these samples due to increasingly fast polarization. Although the proton mobility on H-PHI is not yet clarified, it is likely that these hydrated ions are mobile within the pore structure[23]. However, these much smaller $H^+$ and probably mobile ions via $H_3O^+$) appear to result in less pronounced interactions on the THz scale (akin to water itself), which has a lower $\varepsilon'$ than all $CN_x$ measured herein. In the $\varepsilon''$ spectra, the 2D structure of PHI appears to enable a stronger response than the 1D melon, and the presence of $K^+$ ions seem to increase the THz response and conductivity (*vide infra*) most strongly (see **Supplementary Note 3 & 4**).

**Humidity dependent THz measurements**: Terahertz spectroscopy is widely used to measure water contents in materials, making use of its high absorbance in the THz regime[25–27]. The complex permittivity of water is distinct and well characterised in the THz regime ($\varepsilon'_{1.5\,THz}$ ≈ 4.1 and $\varepsilon''_{1.5\,THz}$ ≈ 1.9)[16,36]. While water has much higher values of $\varepsilon''$ than our materials, its $\varepsilon'$ is smaller than that measured for $CN_x$ in ambient conditions, but higher than that of many common polymers like the ones reported above. Thanks to this difference in $\varepsilon'$ with respect to $CN_x$, the impact of adsorbed or structural water in these materials can also be resolved at THz frequencies. This makes THz-TDS measurements highly relevant for studying $CN_x$, as the practical applications of these materials typically take place in aqueous conditions. Thus, we further measured the studied materials not only in their ambient state, but also when vacuum-dried and at 100% relative humidity, to generate more intrinsic and application relevant insights.

As evidenced from density measurements (**Figure S3**), the $CN_x$ samples studied herein are susceptible to water uptake, with melon #1 changing in density more strongly and taking longer to stabilise than K-PHI # 1 or H-PHI upon drying (ambient density: 1.88 g.cm$^{-3}$; dry: -8.5% for melon, -2.1% for K-PHI). This hydrophilic behaviour enabling water uptake is also qualitatively confirmed by contact angle measurements, with the melon samples being partially wetting (contact angles between 24° and 59°), and K-PHI & H-PHI being fully wetting (**Figure S5**).

**Table 1**: Summary of complex permittivity obtained for hydrophilic materials at 1.5 THz, with red text in brackets indicating percentage change from *ambient* conditions. Fully wetting materials show contact angles less than 20°.

| Material | Dried | | Ambient | | Humid | | Contact Angle |
|---|---|---|---|---|---|---|---|
| | ε' | ε'' | ε' | ε'' | ε' | ε'' | (°) |
| PEG - 8000 | 2.63±0.05 (-5%) | 0.22±0.07 (-16%) | 2.77± 0.01 | 0.19± 0.01 | 2.82 ± 0.03 (+2%) | 0.25 ± 0.04 (+32%) | <20 |
| melon # 1 | 5.46±0.02 (-0%) | 0.26±0.05 (-26%) | 5.60±0.03 | 0.35±0.05 | 5.61±0.02 (+0%) | 0.44± 0.03 (+26%) | 41 ± 2.8 |
| K-PHI | 3.99±0.05 (-38%) | 0.91±0.05 (-44%) | 6.51±0.01 | 1.62±0.02 | 8.24±0.03 (+27%) | 1.56±0.05 (+0%) | <20 |
| H-PHI | 6.33±0.43 (-6%) | 0.57±0.04 (-9%) | 6.00±0.42 | 0.86±0.16 | 6.17±0.19 (+3%) | 0.81±0.25 (-6%) | <20 |

The full (dry, ambient, humid) THz-TDS datasets for melon, K-PHI and H-PHI are shown in **Figure 2**. For melon # 1 (second row, contact angle of 41°) the $\varepsilon'$ dispersion profiles vary little as the sample is vacuum dried or exposed to 100% relative humidity (RH) for 4h, whereas the imaginary permittivity $\varepsilon''$ fluctuates slightly (**Figure 2 d, e** respectively). Similar trends are observed for the other samples melon #2 and melon #3 (**Figure S6**), which are also slightly wetting, but with different contact angles (**Figure S5 a, c, d**). The THz conductivity of melon changes accordingly to its $\varepsilon''$ (**Figure 2f**), i.e. dropping when dried.

For K-PHI on the other hand, which is fully wetting (**Figure S5b**), the water content has a strong influence on the real permittivity (**Figure 2g** and **Table 1**), yielding an increase change of more than 100% from dry ($\varepsilon'_{1.5\ THz}$ = 3.99) to humid ($\varepsilon'_{1.5\ THz}$ = 8.24). The behaviour of $\varepsilon''$ (**Figure 2h**) is more complex with the ambient and wet samples showing similar $\varepsilon''$ above 1 THz, and hence conductivity $\sigma_{THz}$, being ~ 50% higher than the dry sample. It suggests that the THz absorption and thus conductivity properties saturate with ambient water content and do not improve further. Below 1 THz, vacuum drying seems to affect the slope of $\varepsilon''$ – a process that is not yet understood – and potentially affected less by measurement sensitivity in this range, as well as the fact that our K-PHI pellets tend to show structural instability when dried under vacuum.

H-PHI shows effects in-between the other two $CN_x$ samples. $\varepsilon'$ is constant with respect to changes in humidity, akin to melon, but with a 25% higher value (**Table 1**), being close to ambient K-PHI (within error, **Figure 2j**). $\varepsilon''$ follows a behaviour like K-PHI, also approximately doubling upon exposure to ambient conditions, and not changing any more at 100% RH (**Figure 2k**). The values for H-PHI are approximately 47% lower than for ambient K-PHI, indicating amplification effects by $K^+$ ions.

**Figures 2m-o** summarize the humidity dependent data for $\varepsilon'$, $\varepsilon''$ and $\sigma$. Overall, the structurally porous 2D PHI materials have elevated $\varepsilon'$ values compared to melon in ambient or wet conditions, and K-PHI's $\varepsilon'$ response significantly depends on the humid environment. The $\varepsilon''$ of melon is constantly and significantly lower than of the PHI materials. It also increases at 100% RH, which may be related to the water loss or uptake observed in density measurements (**Fig. S3**) and related proton mobility. Both H-PHI and K-PHI approximately double their $\varepsilon''$ upon exposure to ambient humidity. K-PHI shows slightly elevated values and stronger dispersion than H-PHI. $\sigma_{THz}$ reflects this $\varepsilon''$ behaviour: melon has moderate conductivity values that are slightly affected by humidity (32-40 S/m at 1.5 THz), being much lower than of the PHIs. The fact that H-PHI has elevated $\sigma_{THz}$ compared to melon, while increasing more strongly from dry to ambient (47 to ~69 S/m), points to two factors: an inherently increased ultrafast complex permittivity response with the 2D molecular backbone forming pore channels, and water playing a different, more pronounced role when adsorbed *inside* of PHI's confined 1D channels that do not dry out fully.[23]. This is even more pronounced in K-PHI, where $\sigma_{THz}$ is larger than in all humid cases of the other materials when being dry (75 S/m, >2x higher than melon), and doubling again in humid conditions (150 S/m), where it is >4x higher melon and >2x than H-PHI.

Since for THz conductivity, literature values are available, where $\sigma_{THz}$ was obtained directly from THz amplitude changes and using thin-film approximations, these can be put into context of other reported materials for further comparison most easily[37]: triazine based carbon nitrides, poly(triazine imide) = PTI, are structurally similar to PHI and can also contain ions in smaller pores. The ion free PTI-IF, and PTI-LiBr show $\sigma$ ~ 7 and 35 S/m at 1.5 THz in ambient conditions, measured in free-standing pellet form [38]. The conductivity on THz timescales is hence also strongly improved by the presence of ions (x3.7 in PTI, compared to ~x2 in H- vs K-PHI), albeit at lower absolute values, being only 10% for PTI-IF vs H-PHI (7 vs. 70 S/m), and 17% in the ionic case (26 vs. 150 S/m for PTI-LiBr vs. K-PHI, albeit with different ions). A recent report on ion-free tuneable porous covalent organic framework (COF) membranes showed $\sigma_{1.5\,THz}$ between of 5.45 and 75.59 S/m for membrane samples denoted TAM-DBD and TBP-TFP respectively [39], hence being overall in the range of $CN_x$, but still lower than ambient K-PHI.

Our findings suggest that hydrated ion mobility contributes to much higher THz dielectric properties than expected – and that $\varepsilon'$, $\varepsilon''$ and $\sigma$ are hence strongly dependent on the environmental conditions the material is probed in (in dry form as neat material, or humid; more akin to application conditions). We can thus also anticipate that processes occurring on the timescale of THz, i.e. the binding of excitons or charge transfer excitons, and the generation of charges, are positively influenced by the real permittivity in K-PHI in terms of solar energy conversion applications, and especially those in the presence of humidity and water. We hence investigated these materials using TAS.

**fs-ps transient absorption spectroscopy (TAS) measurements:** The real THz permittivity, which corresponds to the picosecond timescale, is an important factor for the Coulomb binding of excitons and the generation of charges by exciton or charge transfer exciton dissociation and their relaxation (**Equation 1**). To provide further evidence of the influence of humidity on the dissociation of (bound) excitons to free charges, we studied the charge carrier generation properties of K-PHI samples in ambient and humid conditions using fs-ps transient pump-probe optical spectroscopy. The results are shown in **Figure 3**.

**Figure 3a and b** show the spectra of a vacuum-dried K-PHI film in ambient and humid conditions from 0.2 ps to 5 ns, with the signal normalized at each delay time to illustrate temporal shape changes (see **Figure S8a,c** for raw data). After excitation with 355 nm light pulses, the material has a pronounced broad absorption signal which appears to red shift from 600 to 670 nm over 5 ns. The latter-timescale 670 nm feature is assigned to electron absorbance in the K-PHI following bandgap excitation[40–44]. From normalized data at each time snapshot, it can be observed that this red-shift begins within the first ps. The shape of the 500-650 nm shoulder is much more pronounced in humid conditions (**Figure 3b**) than in dry conditions (Figure 3a), clearly indicating the impact of humidity on ultrafast dynamics of K-PHI. Using global analysis, or simple subtraction of early (sub-ps) and late (ns) signals, we can deconvolute the spectra into two distinct parts (**Figure 3c**), and compared their amplitude and decay (**Figures S8-S10**) for ambient and humid conditions, respectively. It is possible that the shoulder in the 500-650 nm region is linked to an excitonic or exciton-related state[43]. Since the true nature cannot be clarified yet, we term it *unrelaxed (pre-charge) signal*. This signal peaks at times <1ps and decays while the charge signal rises (peaking at 30-50 ps), consistent with a typical (charge transfer) exciton dissociation into charges, or simply charge related relaxation process. The unrelaxed signal's maximum increases linearly with laser intensity (50-250 µJ/cm², Fig. S8, S9), thereby excluding bimolecular recombination effects. We find that this signal is consistently more pronounced in the wet samples comparted to the ambient ones (~+20%, **Fig. S8e**). Our observation suggests that the more pronounced unrelaxed pre-charge feature existing on the sub-ps timescale is a consequence of water-induced backbone polarization increasing the dielectric screening and hence impacting the photophysics and decay of excited states. Although not being a smoking gun experiment for spectral assignment, this measurement underlines an important role of humidity for modifying the K-PHI real permittivity response to photoexcitation in the 0.5 to 5 ns regime corresponding to the THz frequency range probed. This analysis also directly shows that the yield of this ps-signal can be modified in the material by the presence of water, which appears to be linked to increasing real permittivity in K-PHI that enables the stabilization of light-induced species that relax into stable charges, and hence not to direct influences of water itself.

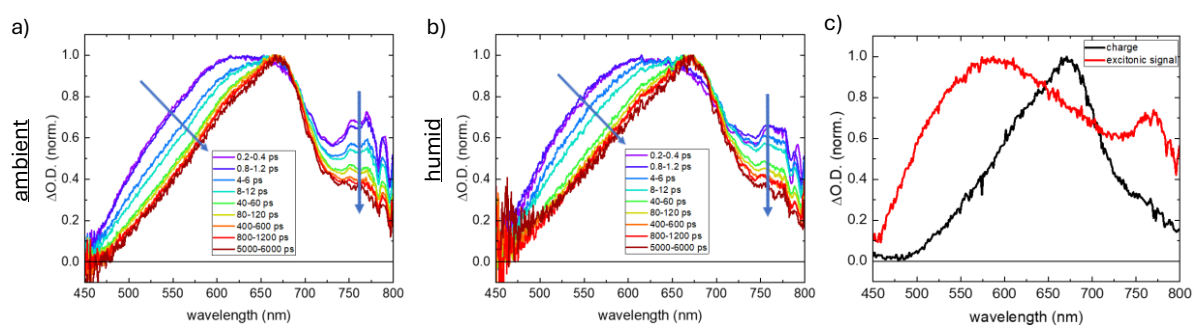

**Figure 3:** fs-ns transient absorption spectra of K-PHI films in **a)** ambient and **b)** humid conditions (>23 hours in water environment), normalized to the maximum signal at each time (fluence: 250 µJ/cm²). Blue arrows indicate regions where a change in spectral shape is observed, mainly due to a more pronounced, unrelaxed pre-charge signal shoulder in humid conditions, centred at 580 nm (see **Figures S8-S10** for details and raw data). **c)** Deconvoluted spectra of the two signals in K-PHI nanosheets: the relaxed charge signal (black) commonly reported, and the unrelaxed, pre-charge signal feature (red), respectively.

**Discussion:** Using THz-TDS, we have shown that the $CN_x$ melon and PHI have a relatively high complex permittivity in the THz regime compared with HDPE, PEG, P3HT and even water, qualifying $CN_x$ as High Refractive Index Polymers (HRIPs), a class of materials that could be useful for many advanced flexible optoelectronic applications such as anti-reflective coatings and encapsulants for OLEDs[45], along with photovoltaic and photocatalytic applications. The real permittivity of proton-terminated melon and H-PHI does not vary much with humidity, but K-PHI shows strong humidity dependence. A direct influence of free water ($\varepsilon'_{1.5\,THz}$ ~4.15), which may be the reason for slight changes in $\varepsilon'$ in presence of humidity (see **Table 1**, PEG e.g.), can be ruled out from PHI's increased permittivity, since the measured $\varepsilon$ spectra cannot be explained by a linear contribution of water to complex permittivity (see **Supplementary Note 5**, **Figure S11**). So far, we can summarize that K-PHI in a dry, intrinsic state has a lower real THz-permittivity than melon, but with the ions present in the 2D structure and responding to water, ε' can be doubled, far beyond melon $CN_x$, H-PHI and water alone (**Figure 2i**).

Influence of structure and ions: In K-PHI, pore ions ($K^+$) are in a hydrated state and loosely bound to the 2D $PHI^-$ backbone[23,24]. We suggest that these ions, confined in 1D pore PHI channels, have an important impact on the photophysics of K-PHI in the THz regime. A closer look shows K-PHI's ambient and humid ε' profile rises slightly initially and drops beyond 0.75 THz (**Figure 2g**). This indicates that the lower frequency range (0.25 – 0.75 THz) might capture a snapshot of the ionic (pore confined, hydrated $K^+$) polarisation contributing initially, and then becoming less active as the $K^+$ ions become unable to respond to the alternating electric field above 1 THz (maximum response time is exceeded). H-PHI on the other hand maintains a relatively constant dispersion profile across the frequency spectrum, possibly due to the much lighter $H^+$ ions having a more constant response across the frequencies probed (**Figure 2j**). The monotonous dispersion trend in H-PHI and its $\varepsilon'$ values close to K-PHI in ambient conditions suggest that the dominant differences between 1D melon and 2D PHI arises from the covalent 2D structure with 1D pore channels in PHI itself, and by the presence of water affecting hydrated ion motion (of $H^+$ and $K^+$) on THz timescales therein. In particular, K-PHI's significantly larger $\varepsilon'$ in humid or aqueous conditions (+100% vs. dry) should in theory faciliate the separation of excitons and charges through decreased Coulomb attraction, whilst more agile polarization should increase charge generation and transport on a fs-ps-timescale. Our fs-TAS data provides preliminary evidence for this, with additional (probably excitonic) signals on the timescale between 0.2 and 100 ps arising more pronouncedly in conditions with higher permittivity.

Relation to complex permittivity and conductivity: Upon drying, a decrease in $\varepsilon'$ is observed in K-PHI, likely related to lower hydrated $K^+$ ion motion due to a reduction in the water content within the 1D pore channel. However, some water is still expected to be trapped in the pores[23,24]. In H-PHI, the residual water may still contribute sufficiently to polarization inducing proton mobility, and thus explain the unchanged $\varepsilon'$ values upon vacuum drying. This can be further rationalized when discussing conductivity, $\sigma_{THz}$ (being linearly related to $\varepsilon''$, **Equation S11**): An earlier report on the dominantly ionic conductivity of metal-containing PHI (*M*-PHI) with different ions, including EIS and NMR probing ionic mobilities on different length scales, indicated that the ionic mobility of *M*-PHI, contributing to the conductivity in the probed $10^{-3}$-$10^{6}$ Hz range, drops significantly when the samples are dried. On the other hand, the enhancement effect of hydrated ion motion showed an onset of saturation at ambient relative humidity (RH) values. This is consistent with our findings, which now extend the frequency regime to the THz timescale: The value of $\varepsilon''_{1.5\,THz}$ the $\sigma_{THz}$ are similar for both ambient and 100% RH conditions for each, H- and K-PHI respectively, and halve in dry conditions. Akin to EIS measurements, the influence of adsorbed structural water on material properties saturates in ambient conditions, and the conductivity effects are stronger with metallic ions. In the EIS regime, H-PHI has conductivities order of magnitude below K-PHI – probably since aqueous protons do not have dominant effects in this mHz-MHz frequency range. In the THz regime, H-PHI shows high conductivity values, but K-PHI has $\sigma_{1.5\,THz}$ 67% higher than H-PHI (see **Figure S4**)**,** indicating that the larger, heavier $K^+$ experience stronger acceleration and resistance within the THz alternating electric field, leading to stronger wave dissipation - but that protons also respond similarly in the 2D PHI network.

The high $\varepsilon'$ values of humid K-PHI cannot be explained solely by the presence of water in isolation. Our humidity data instead suggests that water affects the $\varepsilon'$ of K-PHI through synergistic interactions with ions - likely caused inside structural pore channels. We suggest that this synergy may occur also in other hydrophilic materials. Such an enhancement in polarizability or $\varepsilon'_{THz}$ (for exciton separation and for charge accumulation inside the material) and in $\varepsilon''_{THz}$ (resulting in good conductivity effects on ultrafast fs-ps timescales) seems a key and highly desirable feature for efficient direct conversion of solar energy to fuels in presence of water, as aqueous conditions can apparently boost $\varepsilon'$ responsible for exciton dissociation and charge accumulation in the inner volume of the material, on timescales at which these processes occur, while enabling better charge transport and extraction on the same timescale. These findings are likely a key factor for the enhanced activity for photocatalytic hydrogen evolution reported for PHI-materials compared to melon[23,46], although the overall photocatalytic process is more complex and contains also other factors that need to be considered for efficiencies[47].

Context to other real permittivity techniques: As explained in the introduction, the dielectric properties of materials are strongly dispersive across the frequency range from 1Hz up to ~$10^{15}$ Hz. Given the novelty of the terahertz dielectric property measurements for carbon nitride polymorphs reported herein, we can now place our results in the context of current literature available on these materials, measured with standard techniques such as EIS and Spectroscopic Ellipsometry (SE). These probe the low-mid frequency ($10^{-3}$ to $10^{6}$ Hz), and very high frequency range accessible by UV-NIR ellipsometry ($10^{14}$ to $10^{16}$ Hz), respectively and provide the DC and ultrafast $\varepsilon'$ values often referred to as extremes, since other data is missing. The trends in real permittivity $\varepsilon'$, shown in **Figure 4a**, illustrate that the THz measurements indeed show intermediate values following the expected trend with frequency: EIS estimated $\varepsilon'$ to have values at ~15 and 7 at $10^{-3}$-$10^{5}$ Hz for K-PHI and melon in ambient conditions, respectively [24,48]. Both drop significantly to ~7 and ~5 at 1.5 THz (-53% and -29%). In the $10^{14}$-$10^{15}$ Hz SE-regime (also ambient conditions), melon stays constant (~5)[49], while K-PHI further decreases to ~3 [50], which corresponds to its dry value at ~2 THz.

The parallel trend between EIS and THz-TDS, with K-PHI retaining its higher dielectric response compared with melon in ambient conditions up to the THz regime, indicates that both EIS and THz-TDS are suitable for a *relative* comparison of materials in their real permittivity response, including ambient & humidity effects. SE may capture contributions from electronic polarisation at ultrafast timescales, as these optical excitations are too fast for atomic or molecular rearrangements in either water or $CN_x$ materials, Consequently, we observe a similarity between THz permittivity in the dry state, and the one extracted from SE in both melon and K-PHI.

This shows that drawing conclusions about intermediate and especially THz processes from EIS & SE measurements alone may be insufficient, and that separate THz analysis is crucial for capturing different phenomena on its timescales. This THz-characterization is even more relevant if materials are involved where illumination generates carriers that are key to function on the picosecond timescale. The drop from $\varepsilon'_{DC}$ (static dielectric response) or EIS to $\varepsilon'_{THz}$ is hard to extrapolate, and may be different even for similar materials, as shown here for melon vs PHI. While SE may enable estimation of the THz permittivity response in dry conditions, the result may be strongly modified in reality - by ambient effects, ionic or other phenomena in the THz regime, and on slower times. As such, the real permittivity at the low or high frequency limit provides insufficient information on the functionality of materials used for energy conversion.

Water and wetting groups: The presence of water, or its inclusion by wetting groups with high static real permittivity $\varepsilon'_{DC}$ (e.g. ethylene glycol)[15,51], was discussed earlier in relation to its ability to boost photocatalytic efficiency of polymers e.g. through enhanced permittivity (measured by EIS) [52–54]. While water has a high dielectric response in the low frequency regime, this is not the case in the THz regime. Hence, it can increase the response of materials with lower $\varepsilon'$ values than itself ($\varepsilon'$ ~ 2.77 for ambient PEG-8,000 vs 4.1 for water at 1.5 THz) from a linear combination of water + material, especially in wetting materials, but this effect may be smaller than expected (2-5% changes observed for poly(ethylene glycol)). However, it was not previously clear a priori that $\varepsilon'$ can be increased for materials that have larger $\varepsilon'$ values than water, like $CN_x$[26], where we now revealed synergistic effects beyond those expected from a linear combination of the material and water alone. Hence, when charge stabilizing effects through wetting groups are observed on in ultrafast (fs-ns) spectroscopy measurements (and not only on sub-µs time scales corresponding to EIS analysis) in solar cell or photocatalyst materials, they may likely be linked also to modified THz properties, which can in future be studied separately by THz-TDS – giving a better approximation to applications also involving water, or other environments in principle.

$\varepsilon'_{THz}$ in context to other materials: **Figure 4b** shows the real permittivity of typical inorganic semiconductors used for solar energy conversion and compares them with organic materials and our study. Indeed, we find slightly elevated values for solar cell materials like Si or GaAs ($\varepsilon'_{1.5THz}$ >10), but many others show only modest values being in the range of humidity variations of K-PHI, i.e. $\varepsilon'_{1.5THz}$=3-8: E.g. the hybrid perovskite $CH_3NH_3PbI_3$, II-VI semiconductors like CdTe, and oxide semiconductors like $TiO_2$ or $Fe_2O_3$. The often-made generalization that inorganic materials have higher dielectric response than organic materials, enabling better exciton separation[4], may therefore not be universally true in the THz regime. We also note that exchange of the mobile A-site cation in the lead halide perovskites from methylammonium ($CH_3NH_3$) to Cs also show drastic changes in THz permittivity ($\varepsilon'_{1.5THz}$ changes from ~7.7[55] to ~20.5[56]) despite only a fraction of the material being modified.

Solar energy conversion: Despite the importance of understanding (modified) permittivity on THz timescales vs. mHz-MHz timescales (EIS) or SE, it remains an open question as to which timescale is most relevant for *enhancing* light-energy conversion function .The result likely depends on the application scenario and on the bottleneck in the process, i.e. on morphological factors (e.g. particle size, gran boundary density and their effects), on the function and density of intrinsic traps, and on transport properties – i.e. the ability to make use of light generated charges after collecting them [4,47,57,58]. While the fast timescales are crucial for exciton dissociation and charge carrier generation, they don't have to be limiting (see hybrid perovskites, or $TiO_2$ enabling 100% photocatalytic quantum efficiency as perfectly engineered cases[59]), and slower (µs-s) times scales may also be highly relevant for photocatalytic function, since surface catalytic processes are often slow, requiring charge stabilization[60,61].

A common method to increase the low frequency permittivity is to add functional groups like ethylene glycol with high $\varepsilon'$ values at low frequences to polymers with intrinsically lower $\varepsilon'$ values[15,51]. These functionalizations are reported to increase the conversion efficiency of sunlight to hydrogen in photocatalysis due to various effects occurring on different timescales including the ns-regime[15], but their influence is most clear in the lifetime of charges extending to the µs-s range (corresponding to EIS frequencies, i.e. 1 Hz-1 MHz) due to enhanced polarizability of the material. However, less is known about the functional groups' direct influence on exciton dissociation to charges, which still is a major bottleneck in photocatalysis with organic materials and related solar energy conversion technologies. THz-TDS complex permittivity analysis, carried out herein, can in principle provide insights on the origin of photophysical interaction effects expected in the sub-ps timescale, and thus directly support description of exciton stability (lifetime) and charge generation. At present, such phenomena are typically measured using fs-ns TAS, which is unable to provide direct photophysical material properties and can be difficult to interpret due to the high fluxes used in the excitation pulses. Further, for the purposes of material modelling and property prediction, knowledge of the complex permittivity at given timescales is key to correctly elucidate corresponding properties and behaviour of functional interest. This is especially key in the context of Artificial Intelligence (AI) and Machine Learning, e.g. if targeting improved synthetic (bottom-up) design, which is particularly promising with organic materials. THz measurements easily provide such valuable insights, without complex models or assumptions being required.

Potential for light-driven ionics: Enhanced THz $\varepsilon'$ relative to melon is herein clearly visible. Further, the presence of aqueous alkali metal ions contribute to THz polarization of the material, and to differently pronounced relaxation mechanisms on the ps-scale. Since permittivity increases because of higher polarizability, the enhanced $\varepsilon'$ in K-PHI indirectly show that light-driven polarization (through charge formation e.g.) may in turn affect structural hydrated $K^+$ ions on picosecond timescales. As such, these insights also possibly point to efficient *optoionic* interactions[62,63] in K-PHI (being mediated by water) occurring on the fs-ps regime - where light generated charge carriers interact with internal or external (solvated) ions, modifying the material's stoichiometry and with that, photophysical properties. In K-PHI, the presence of these alkali metal ions enables photocharging, i.e. the intrinsic combination of light absorption (solar cell function) and charge storage (battery function) properties. Besides photobatteries, this can be used for effects like dark photocatalysis, for powering microparticles, and for sensing etc[40–42,50,64]. An ionically enhanced permittivity on the THz timescale is expected to be beneficial for generating and/or stabilizing photogenerated charges quickly and efficiently. We therefore suggest that the overall strength optoionic effects and the efficiency of photocharging materials and light-driven ionotronic devices may benefit from enhanced THz permittivities[65–67].

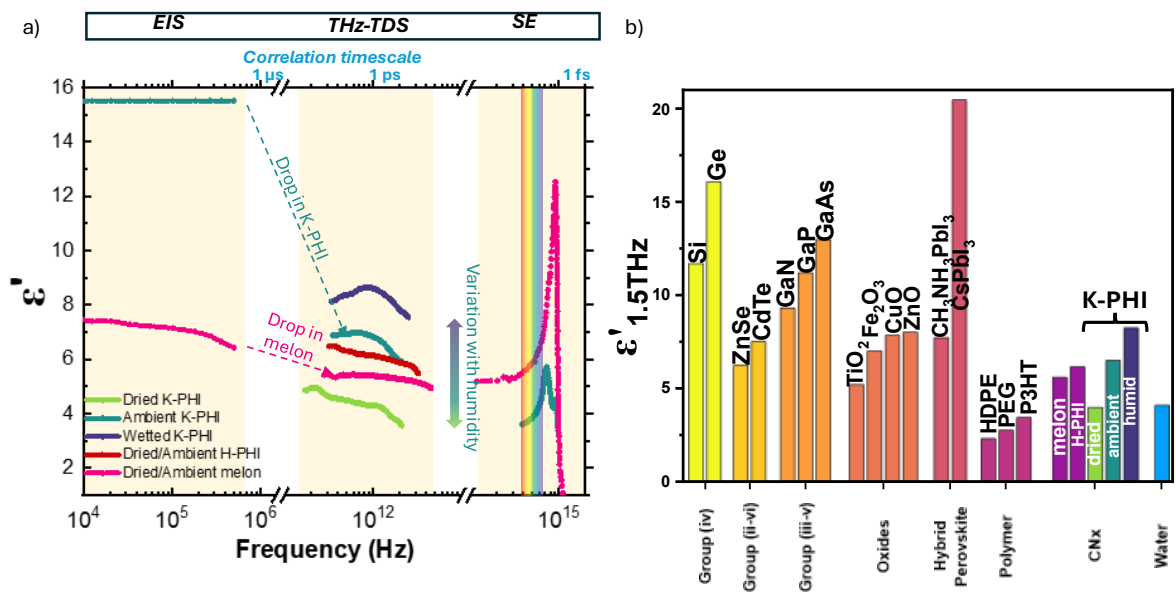

**Figure 4**: **a)** Real permittivity comparison over the frequency spectrum for melon and K-PHI, with data extracted from Electrochemical Impedance Spectroscopy (EIS), THz (data from **Figure 2**) and Spectroscopic Ellipsometry (SE) (literature values reproduced under CC by 4.0 license)[49,50,68], confirming the predicted dispersive trends in real permittivity (see **Figure 1a**). **b)** Real Permittivity at 1.5 THz for common semiconductor materials used for solar energy conversion applications, relative to measured $CN_x$ and water[55,56,77,69–76].

**Summary & outlook:**

In this work, we have shown that THz-TDS is a powerful, non-destructive, and non-contact tool for directly characterizing the complex permittivity of organic semiconductors. It provides valuable insights into their dielectric response and related properties (e.g. THz conductivity). This characterization on the (sub-)ps-timescale closes the "THz gap" (**Figure 1a**) resulting from limitations to EIS (mHz to $10^6$ Hz), microwave dielectric measurements (GHz timescales) and SE ($10^{14}$ to $10^{16}$ Hz). Different polarization processes occur on respective timescales, resulting in timescale-dependent trends and environmental effects that are difficult to extrapolate based on measurements made on a different timescale. The THz regime encompasses processes on time scales involving exciton generation and separation as well as fast charge relaxation effects. Hence, the characterization of complex permittivity by THz-TDS can provide a broader understanding of functional materials used for light conversion applications, including photovoltaics, photodetectors and photocatalysts, and enable more accurate modelling from knowledge of (photo)physical properties. For example, **Figure 4b** compares the $\varepsilon'_{1.5THz}$ of materials used for such applications, pointing out that inorganics used for solar energy conversion are not necessarily high dielectrics in the THz regime, and that $CN_x$ has comparable dielectric properties on this timescale.

We further highlight the importance of structurally adsorbed water being able to synergistically enhance the complex permittivity in the THz-regime. Here, a more straightforward and unambiguous characterization of contributions from the backbone, water and ions become possible when comparing to probing techniques like EIS and SE. Hence, care must be taken when comparing measurements acquired in different conditions, and when materials are employed in humid environments, it is important bear in mind that the properties of dry materials may be significantly different from those occurring in real application scenarios (such as photocatalytic reactions in water or other electrolytes). This is also relevant for modelling purposes, which often neglect environmental factors, or might use dielectric properties as input parameters, which cannot be generalized from other frequency domains, and would neglect synergistic effects seen here in the THz regime. The influence of increased humidity on the excitonic signals in K-PHI was further evidenced by fs-ps-TAS (**Fig. 3**), pointing to stabilization of the former.

The use of THz-TDS, being increasingly available, it is hence highly recommended for characterization of functional materials used in solar energy conversion, or with crucial function in the (sub-)ps regime. In particular, the surprising findings of structural water or hydrated ions apparently contributing to the real dielectric permittivity response even on ps timescales (promoted via ambient humidity) sheds further light on material analysis and selection for future *ionotronic* materials operating on THz timescales – for which K-PHI or other porous framework materials containing ions, like COFs, MOFs or porous ionic polymers may be good candidates. THz-TDS enables direct extraction of their conductivity $\sigma_{THz}$ without further approximations. Also, direct light-driven ionic effects (*optoionics*) may potentially be characterized more directly by this technique in the future. The ability of soft ionic materials to exhibit strongly enhanced permittivities in the presence of water may help to combat the often-limiting strong exciton recombination seen in apparently low dielectric materials – and often targeted by wetting – although without knowledge of fs-ps time scale influence. THz-TDS may hence be a key component for addressing fundamental challenges of organic semiconductors and their applicability for solar energy conversion, and thus be an enabler in the transition to a more sustainable energy economy and the development of related materials infrastructure.

**Acknowledgements:**

The authors acknowledge Dr. Carolina Pulignani and Prof. Erwin Reisner (Dep. Of Chemistry, University of Cambridge) for providing carbon nitrides for this study, Dr. Nuria Tapia-Ruiz and Sarah McKinney for use of their labs and assistance in pellet preparation, and Dr. Andrew Gregory for use of his in-house NPL THz-TDS software for thickness optimisation. Dr. Julia Kröger (EnBW) and Prof. Ji-Seon Kim (Imperial) are acknowledged for fruitful discussions. F.P. and J.D. acknowledge funding from UKRI (grant reference EP/X027449/1) and EPSRC (grant reference EP/T028513/1). P. A. thanks the Science and Solutions for a Changing Planet DTP (NE/S007415/1), hosted at the Grantham Institute for Climate Change and the Environment for funding this research. A. K. thanks the EPSRC for a Programme Grant (EP/W017075/1). Work at the National Physical Laboratory was supported by the UK Government's Department for Science, Innovation and Technology (DSIT) through the UK's National Measurement System programs.

**Author contributions:**

J.D., M.N and F.P. designed and supervised the project. F.P. and P.A. synthesized and prepared the $CN_x$ materials. D.S. and S.S. performed and analysed density measurements, P.A the XRD measurements. R.J. and F.P prepared $CN_x$ pellets for characterisation. R.J. and F.P. measured contact angles. R.J.

measured and analysed the THz data, with assistance and supervision of M.N. F.P. and S.H. prepared, performed, and analysed fs-ns TAS measurements. F.P. and R.J wrote the manuscript draft. All authors incl. A.K. contributed to manuscript revisions and discussions.

**Conflict of interest:**

The authors declare no conflict of interest.

**Data availability:**

The data reported in this study will be shared on the research repository *zenodo* following acceptance.